\documentclass[twocolumn,showpacs,aps,prl,superscriptaddress]{revtex4}

\usepackage{graphicx}
\usepackage{dcolumn}
\usepackage{amsmath}
\usepackage{epsfig}

\input pubboard/babarsym

\RequirePackage{xspace}

\newcommand{\pvec}{{\bf p}}

\newcommand{\acp}{\ensuremath{\calA_{ch}}}
\newcommand{\calB}{\ensuremath{{\cal B}}}

\newcommand{\DE}{\ensuremath{\Delta E}}
\newcommand{\mb}{\ensuremath{m_{\rm ES}}}
\newcommand{\mres}{\ensuremath{m_{\rm res}}}
\newcommand{\xf}{\ensuremath{{\cal F}}}
\newcommand{\hel}{\ensuremath{{\cal H}}}
\newcommand{\thetaT}{\ensuremath{\theta_{\rm T}}}
\newcommand{\costhr}{\ensuremath{\cos\thetaT}}

\newcommand\etal{{\it et al.}}
\newcommand{\half}{\ensuremath{{1\over2}}}

\newcommand{\bma}[1]{\boldmath{$#1$}}

\newcommand{\bfig}{\begin{figure}[htbpc!]}
\newcommand{\efig}{\end{figure}}
\newcommand\bef{\begin{figure}}
\newcommand\edf{\end{figure}}
\newcommand\dbline{\noalign{\vskip 0.10truecm\hrule}\noalign{\vskip 2pt}\noalign{\hrule\vskip 0.10truecm}}
\providecommand{\tbline}{\noalign{\vskip 0.05truecm\hrule\vskip0.05truecm}}

\newcommand\beq{\begin{equation}}
\newcommand\eeq{\end{equation}}
\newcommand\bear{\begin{array}}
\newcommand\enar{\end{array}}
\newcommand\beqa{\begin{eqnarray}}
\newcommand\eeqa{\end{eqnarray}}
\newcommand\ben{\begin{enumerate}}
\newcommand\een{\end{enumerate}}

\newcommand{\UfourS}{\ensuremath{\Upsilon(4S)}}

\newcommand{\etagg}{\ensuremath{\eta_{\gaga}}}
\newcommand{\etappp}{\ensuremath{\eta_{3\pi}}}
\newcommand{\etatogg}{\ensuremath{\eta\ra\gaga}}
\newcommand{\etatoppp}{\ensuremath{\eta\ra\pi^+\pi^-\pi^0}}

\newcommand{\etapepp}{\ensuremath{\etapr_{\eta\pi\pi}}}

\newcommand{\etaptoeggpp}{\ensuremath{\etapr\ra\etagg\pip\pim}}
\newcommand{\etaprg}{\ensuremath{\etapr_{\rho\gamma}}}
\newcommand{\etaptorg}{\ensuremath{\etapr\ra\rho^0\gamma}}

\newcommand{\omtoppp}{\ensuremath{{\omega\ra\pip\pim\piz}}}

   \newcommand{\rhop}{\ensuremath{\rho^+}}
   
\newcommand{\kzs}{\ensuremath{\KS}}

\newcommand{\fetapip}{\ensuremath{\eta\pi^+}}
\newcommand{\etapip}{\ensuremath{\Bp\ra\fetapip}}
\newcommand{\Betapip}{\ensuremath{\calB(\etapip)}}
\newcommand{\retapip}{\ensuremath{xx^{+xx}_{-xx}\pm xx}}

\newcommand{\Aetapip}{\ensuremath{xx^{+xx}_{-xx}\pm xx}}

\newcommand{\setapip}{\ensuremath{xx}}

   \newcommand{\fetaggpip}{\ensuremath{\eta_{\gaga} \pip}}
   
   \newcommand{\fetappppip}{\ensuremath{\eta_{3\pi} \pip}}

\newcommand{\fetaKp}{\ensuremath{\eta K^+}}
\newcommand{\etaKp}{\ensuremath{\Bp\ra\fetaKp}}
\newcommand{\BetaKp}{\ensuremath{\calB(\etaKp)}}
\newcommand{\retaKp}{\ensuremath{xx^{+xx}_{-xx}\pm xx}}

\newcommand{\AetaKp}{\ensuremath{xx^{+xx}_{-xx}\pm xx}}

\newcommand{\setaKp}{\ensuremath{xx}}

   \newcommand{\fetaggKp}{\ensuremath{\eta_{\gaga} \Kp}}
   
   \newcommand{\fetapppKp}{\ensuremath{\eta_{3\pi} \Kp}}

\newcommand{\fetaKz}{\ensuremath{\eta\Kz}}
\newcommand{\etaKz}{\ensuremath{\Bz\ra\fetaKz}}
\newcommand{\BetaKz}{\ensuremath{\calB(\etaKz)}}
\newcommand{\retaKz}{\ensuremath{xx^{+xx}_{-xx}\pm xx}}

\newcommand{\uletaKz}{\ensuremath{xx}}

\newcommand{\setaKz}{\ensuremath{xx}}

   \newcommand{\fetaggks}{\ensuremath{\eta_{\gaga}\KS}}
   
   \newcommand{\fetaggKz}{\ensuremath{\eta_{\gaga}\Kz}}
   
   \newcommand{\fetapppKz}{\ensuremath{\eta_{3\pi}\Kz}}

\newcommand{\fetaomega}{\ensuremath{\eta\omega}\xspace}
\newcommand{\etaomega}{\ensuremath{\Bz\ra\fetaomega}\xspace}
\newcommand{\Betaomega}{\ensuremath{\calB(\etaomega)}\xspace}
\newcommand{\retaomega}{\ensuremath{xx^{+xx}_{-xx}\pm xx}\xspace}

\newcommand{\uletaomega}{\ensuremath{xx}\xspace}

\newcommand{\setaomega}{\ensuremath{xx}\xspace}
   \newcommand{\fetaggomega}{\ensuremath{\eta_{\gaga} \omega}\xspace}
   
   \newcommand{\fetapppomega}{\ensuremath{\eta_{3\pi} \omega}\xspace}

\newcommand{\fetarhop}{\ensuremath{\eta\rho^+}}
\newcommand{\etarhop}{\ensuremath{\Bp\ra\fetarhop}}
\newcommand{\Betarhop}{\ensuremath{\calB(\etarhop)}}
\newcommand{\retarhop}{\ensuremath{xx^{+xx}_{-xx}\pm xx}}

\newcommand{\Aetarhop}{\ensuremath{xx\pm xx \pm xx}}
\newcommand{\setarhop}{\ensuremath{xx}}
  \newcommand{\fetaggrhop}{\ensuremath{\eta_{\gamma\gamma} \rho^+}}
  
  \newcommand{\fetappprhop}{\ensuremath{\eta_{3\pi} \rho^+}}

\newcommand{\fetappip}{\ensuremath{\etapr\pip}}
\newcommand{\etappip}{\ensuremath{\Bp\ra\fetappip}}
\newcommand{\Betappip}{\ensuremath{\calB(\Bp\ra\etapr \pip)}}
\newcommand{\retappip}{\ensuremath{xx^{+xx}_{-xx} \pm xx}}

\newcommand{\Aetappip}{\ensuremath{xx\pm xx}}
\newcommand{\setappip}{\ensuremath{xx}}
   \newcommand{\fetapepppip}{\ensuremath{\etapr_{\eta\pi\pi} \pi^+}}
   
   \newcommand{\fetaprgpip}{\ensuremath{\etapr_{\rho\gamma} \pi^+}}
   \newcommand{\etaprgpip}{\ensuremath{\Bp\ra\fetaprgpip}}

\newcommand{\BABARPubYear}    {05}
\newcommand{\BABARPubNumber}  {003}

\newcommand{\SLACPubNumber} {11079}

\renewcommand{\retaomega}{\ensuremath{1.0\pm 0.5\pm 0.2}}
\renewcommand{\uletaomega}{\ensuremath{1.9}}
\renewcommand{\setaomega}{\ensuremath{2.5}}

\renewcommand{\retaKz}{\ensuremath{1.5\pm0.7\pm 0.1}}
\renewcommand{\uletaKz}{\ensuremath{2.5}}
\renewcommand{\setaKz}{\ensuremath{2.6}}

\renewcommand{\retarhop}{\ensuremath{8.4\pm 1.9\pm 1.1}}

\renewcommand{\Aetarhop}{\ensuremath{0.02\pm 0.18 \pm 0.02}}
\renewcommand{\setarhop}{\ensuremath{4.7}}

\renewcommand{\retappip}{\ensuremath{4.0\pm 0.8\pm 0.4}}

\renewcommand{\setappip}{\ensuremath{5.4}}
\renewcommand{\Aetappip}{\ensuremath{0.14\pm0.16\pm 0.01}}

\renewcommand{\retapip}{\ensuremath{5.1\pm0.6\pm 0.3}}
\renewcommand{\setapip}{\ensuremath{9.7}}
\renewcommand{\Aetapip}{\ensuremath{-0.13\pm 0.12\pm 0.01}}

\renewcommand{\retaKp}{\ensuremath{3.3\pm0.6\pm 0.3}}
\renewcommand{\setaKp}{\ensuremath{6.7}}
\renewcommand{\AetaKp}{\ensuremath{-0.20\pm 0.15\pm 0.01}}

\def\figurebox#1#2#3{%
    \def\arg{#3}%
    \ifx\arg\empty
    {\hfill\vbox{\hsize#2\hrule\hbox to #2{\vrule\hfill\vbox to #1{\hsize#2\vfill}\vrule}\hrule}\hfill}%
    \else
    {\hfill\epsfbox{#3}\hfill}%
    \fi}

\begin{document}

\preprint{\babar-PUB-\BABARPubYear/\BABARPubNumber} 
\preprint{SLAC-PUB-\SLACPubNumber} 

\begin{flushleft}
\babar-PUB-\BABARPubYear/\BABARPubNumber\\
SLAC-PUB-\SLACPubNumber\\
\end{flushleft}

\title{
\large \bf \boldmath
Measurement of branching fractions and charge asymmetries in $B^+$ decays to
\fetapip , \fetaKp , \fetarhop\ and \fetappip , and search for $B^0$ decays to
\fetaKz\ and \fetaomega }

%
\author{B.~Aubert}
\author{R.~Barate}
\author{D.~Boutigny}
\author{F.~Couderc}
\author{Y.~Karyotakis}
\author{J.~P.~Lees}
\author{V.~Poireau}
\author{V.~Tisserand}
\author{A.~Zghiche}
\affiliation{Laboratoire de Physique des Particules, F-74941 Annecy-le-Vieux, France }
\author{E.~Grauges-Pous}
\affiliation{IFAE, Universitat Autonoma de Barcelona, E-08193 Bellaterra, Barcelona, Spain }
% ...
\collaboration{The \babar\ Collaboration}
\noaffiliation

\date{\today}

\begin{abstract}
We present measurements of branching fractions and charge asymmetries for six $B$-meson
decay modes with an $\eta$ or \etapr\ meson in the final state. The data sample corresponds
to 232 million \BB\ pairs collected with the \babar\ detector 
at the PEP-II asymmetric-energy \epem\ $B$ Factory at SLAC.
We measure the branching fractions (in units of $10^{-6}$):
$\Betapip=\retapip$,
$\BetaKp=\retaKp$,
$\BetaKz=\retaKz$ ($<\uletaKz$ at 90\% C.L.),
$\Betarhop=\retarhop$,
$\Betaomega=\retaomega$ ($<\uletaomega$ at 90\% C.L.),
and $\Betappip=\retappip$,
where the first uncertainty is statistical and second systematic.
We also determine the charge asymmetries for the charged modes:
$\acp(\etapip)=\Aetapip$,
$\acp(\etaKp)=\AetaKp$,
$\acp(\etarhop)=\Aetarhop$, and
$\acp(\etappip)=\Aetappip$.
\end{abstract}


\maketitle


Charmless $B$ decays are becoming increasingly useful to test the
accuracy of theoretical predictions,
for example based on QCD factorization~\cite{SUthreeQCDFact,acpQCDfact} or
flavor~SU(3) symmetry~\cite{FUglob,chiangGlob}.
In this Letter we present measurements
of branching fractions and, when applicable, charge asymmetries for six charmless $B$ decays:
\etapip, \etaKp, \etaKz, \etaomega, \etarhop\ and \etappip,
of which the last four were not observed before~\cite{PRLetah,PRD,isoscalarPRL,BellePRD}.
Some of these decays may proceed through CKM-suppressed $b\ra u$ and
loop (``penguin") $b\ra s$ transitions with amplitudes of comparable size.
Interference between these amplitudes can lead to direct \CP\ violation
measurable in charge asymmetries \acp~\cite{acpQCDfact}. The measured branching fractions
and charge asymmetries may also be sensitive to the effect of non-Standard-model heavy particles entering
the loop~\cite{beyondSM}.

Charmless $B$ decays with kaons are usually expected to be dominated by
$b\ra s$ loop amplitudes, while $b\ra u$ tree amplitudes
are typically larger for the decays with pions and $\rho$ mesons.
However, the $\B\ra\eta K$ decays are especially 
interesting since they are suppressed relative to the abundant $\B\ra\etapr K$ 
decays due to destructive interference between two penguin amplitudes
\cite{Lipkin}.
The CKM-suppressed $b\ra u$ tree amplitudes may interfere significantly with
$b\ra s$ penguin amplitudes of similar sizes, possibly leading to large direct \CP\ violation in
\etarhop, \etapip\ and \etappip \cite{directCP}; numerical estimates are
available in a few cases \cite{acpQCDfact,FUglob,acpgrabbag}.  We search for
such direct  
\CP\ violation by measuring the charge asymmetry $\acp \equiv
(\Gamma^--\Gamma^+)/(\Gamma^-+\Gamma^+)$ in the rates
$\Gamma^\pm=\Gamma(B^\pm\ra f^\pm)$ for each
charged final state $f^\pm$.

Finally, phenomenological fits to the branching fractions and charge asymmetries
of charmless $B$ decays
can be used to understand the relative importance of tree and penguin contributions 
and may provide sensitivity to the CKM angle $\gamma$~\cite{FUglob,chiangGlob}.


The results presented here are obtained from extended unbinned maximum
likelihood (ML) fits to data collected with the \babar\
detector~\cite{BABARNIM} at the PEP-II asymmetric \epem\
collider~\cite{pep} located at the Stanford Linear Accelerator Center.
The analysis uses an integrated luminosity of 211~fb$^{-1}$, corresponding
to $232$ million \BB\ pairs, recorded
at the $\Upsilon (4S)$ resonance (center-of-mass energy $\sqrt{s}=10.58\
\gev$).


Charged particles
are detected and their
momenta measured by a combination of a vertex tracker consisting
of five layers of double-sided silicon microstrip detectors and a
40-layer central drift chamber, both operating in the 1.5-T magnetic
field of a superconducting solenoid. We identify photons and electrons 
using a CsI(Tl) electromagnetic calorimeter (EMC).
Charged particle identification (PID) is provided by
an internally reflecting ring imaging Cherenkov detector (DIRC)
covering the central region of the detector,
the average energy loss ($dE/dx$) in the tracking devices and by
the EMC.
A $K/\pi$ separation better than two standard deviations
($\sigma$) is achieved for all momenta.


We select $\eta$, $\etapr$, $\omega$, \kzs\ and $\piz$ candidates through the decays
\etatogg\ (\etagg), \etatoppp\ (\etappp), 
\etaptoeggpp\ (\etapepp), \etaptorg\ (\etaprg),
\omtoppp,
$\kzs\ra\pip\pim$ and $\piz\ra\gaga$.  We
impose the following requirements on the invariant mass in \mev\ of the particle candidates' final
states: $490< m_{\gaga}<600$ for \etagg, $520<m_{\pi\pi\pi}< 570$ for
\etappp, $910<(m_{\eta\pi\pi},m_{\rho\gamma})<1000$ for \etapr,
$735<m_{\pi\pi\pi}<825$ for $\omega$, 
$510 < m_{\pi\pi} <1070$ for $\rho^0$, $470 < m_{\pi\pi} <1070$ for $\rho^+$,
$486 < m_{\pi\pi} < 510$ for \kzs\
and $120 < m_{\gaga} < 150$ for \piz.
These cuts are loose for the invariant mass variables used in the ML fit,
and tight for those that are not.
For \kzs\ candidates we require at least $3\sigma$ three-dimensional separation between
the decay vertex and the \epem\ collision point.
For the vector resonances $\omega$ and \rhop\ we also use the
helicity-frame decay angle $\theta_H$.
The helicity frame is defined as the vector-meson rest frame
with polar axis along the direction of the boost from the $B$ rest
frame.  For $\omega$, $\theta_H$ is the polar angle of the normal to the
decay plane, and for $\rho$ it is the polar angle of the charged daughter
momentum. 
We define $\hel\equiv \cos{\theta_H}$ and require $-0.75<\hel<0.95$ for
\rhop.

All tracks from resonance candidates are required to have PID
consistent with pions. For the \Bp\ decays
to \fetapip, \fetaKp\ and \fetappip,
the primary charged 
track must have an associated DIRC Cherenkov angle within $3.5\sigma$
of the expected value for either a $\pi$ or $K$ hypothesis.
The discrimination between primary $\pi$ and $K$ is performed
in the ML\ fits.

A $B$-meson candidate is characterized kinematically by the
energy-substituted mass
$\mes=(\frac{1}{4}s-\pvec_B^2)^\half$
and energy difference $\DE = E_B-\half\sqrt{s}$, where
$(E_B,\pvec_B)$ is the $B$-meson 4-momentum vector, and
all values are expressed in the \UfourS\ frame.
Signal events peak at zero for \DE , and at the $B$ nominal mass for \mes .
The resolution on \DE\ (\mes) is about 30 MeV ($3.0\ \mev$).
We require $|\DE|\le0.2$ GeV and $5.25\le\mes\le5.29\ \gev$.

Backgrounds arise primarily from random combinations in continuum $\epem\ra\qqbar$
($q=u,d,s,c$) events. To reject these events we make use of the angle
\thetaT\ between the thrust axis of the $B$ candidate in the \UfourS\
frame and that of the 
rest of the charged tracks and neutral clusters in the event.
The distribution of $|\costhr|$ is
sharply peaked near $1$ for combinations drawn from jet-like \qqbar\
pairs, and nearly uniform for the almost isotropic $B$-meson decays; we
require $|\costhr|<0.9$ ($<0.65$ for \fetaprgpip). Further discrimination
from continuum in the ML\ fit is obtained from a Fisher discriminant \xf\
that is described in detail elsewhere~\cite{PRD}.

Where necessary, we use additional event selection criteria to reduce \BB\ backgrounds from
several charmless final states. Specifically, we require that photons
have energies 
in ranges uncharacteristic of these backgrounds and, in \fetaggomega\ and
\fetaggks, we eliminate \etagg\ candidates that share a photon with any \piz\ candidate
having momentum between 1.9 and 3.1 GeV in the \UfourS\ frame. 

Multiple candidates are found in less than 30\% of the events, in which case we choose the candidate with
the smallest value of a $\chi^2$ constructed from the deviations of the
daughter resonance masses from their nominal values.


We use Monte Carlo (MC) simulation \cite{geant} for an initial estimate
of the residual \BB\ background and
to identify the few (mostly charmless) decays that may
survive the candidate selection and have characteristics similar to the signal.
We find these contributions to be negligible for several of our modes.
Where they are not negligible, namely for \fetaggpip, \fetaggKp, \fetarhop\ and \fetaprgpip,
we include a component in the ML fit to account for them.


We obtain yields and \acp\ for each decay chain from a
ML fit with the following input observables: \DE, \mes, \xf, and
$\mres$ (the mass of the $\eta$, \etapr, $\rho^+$, or $\omega$ candidate).
For $\omega$ and \rhop\ decays we also use
\hel\ and, for charged modes with a primary charged track,
the PID variables $S_{\pi}$ and $S_K$, defined as
the number of standard deviations between the measured DIRC Cherenkov angle
and that expected for pions and kaons, respectively.

For each event $i$, hypothesis $j$ (signal, continuum background, 
\BB\ background), and flavor $k$ (primary \pip\ or \Kp),
we define the  probability density function (PDF)
\begin{eqnarray}
{\cal P}^i_{jk} =  {\cal P}_j (\mes^i) {\cal  P}_j (\DE^i_k\left[,S^i_k\right]) 
 { \cal P}_j(\xf^i) {\cal P}_j (\mres^i\left[,\hel^i\right])
\end{eqnarray}
The bracketed variables $S$ and \hel\ pertain to modes with a primary charged
track or vector resonance daughters, respectively.
Known correlations between $\DE_k$ and $S_k$, and between \mres\ and \hel ,
are included in the PDF.
The likelihood function is
\begin{equation}
{\cal L} = \exp{(-\sum_{j,k} Y_{jk})}
\prod_i^{N}\left[\sum_{j,k} Y_{jk} {\cal P}^i_{jk}\right]\,,
\end{equation}
where $Y_{jk}$ is the yield of events of hypothesis $j$ and flavor $k$,
to be found by maximizing \calL . $N$ is the number of events in the sample.
Free parameters of the fit are the signal and background yields,
\qqbar\ background PDF parameters (see below), and for charged modes the signal and \qqbar\
background charge asymmetries.


For the signal and \BB\ background components we determine the PDF
parameters from simulation.
For background from continuum (and non-peaking combinations from $B$ decays)
we obtain the PDF from (\mb,\,\DE) sideband data for each decay chain,
before applying the fit
to data in the signal region; we refine this PDF by letting as many of its
parameters as feasible free to vary in the final fit.
We parameterize each of the functions ${\cal P}_{\rm
sig}(\mes),\ {\cal P}_{\rm sig}(\DE_k),\ { \cal P}_j(\xf),\ { \cal
P}(S_k)$ and the peaking components of ${\cal P}_j(\mres)$ with either
a Gaussian, the sum of two Gaussians or an asymmetric Gaussian function
as required to describe the distribution.  Slowly varying distributions
(mass, energy or helicity-angle for combinatorial background) are
represented by one or a combination of linear, quadratic and phase-space
motivated functions~\cite{PRD}.
The peaking and combinatorial components of the $\omega$ and \rhop\
mass spectra each have their own $\hel$ shapes.
Control samples with similar topologies as our signal modes (e.g.\ $B\ra D(K\pi\pi)\pi$) are used
to verify or adjust the simulated resolutions evaluated from MC.


Before applying the fitting procedure to the data we subject it to
several tests.
In particular we evaluate possible biases in the yields from our neglect of
small residual
correlations among 
discriminating variables in the PDFs. This is achieved by fitting ensembles of simulated
\qqbar\ experiments drawn from the PDF into which we have embedded the expected
number of signal and \BB\ background events, randomly extracted from the fully simulated MC
samples. The measured biases are listed in Table~\ref{tab:results}.


\begin{table*}[btp]
\caption{
Fitted signal yield $Y_S$ in events (ev.), estimated purity $P$, measured bias (see text), detection
efficiency $\epsilon$, daughter branching fraction product ($\prod\calB_i$),
significance~$\cal S$ (with systematic uncertainties included), measured branching fraction \calB ,
and signal charge asymmetry \acp\ for each mode.
The quantities in parentheses are 90\% C.L. upper limits.
}
\label{tab:results}
\begin{tabular}{lcccccccc}
\dbline
Mode	      	& $Y_S$	(ev.)   		
					&$P$ (\%)
						& Bias (ev.)
							&$\epsilon$ (\%)
								&$\prod\calB_i$ (\%)
									& $\cal S$ ($\sigma$)	&  \calB\ $(10^{-6})$	& \acp\			\\
\tbline
~~\fetaggpip	&$153^{+24}_{-22}$	
					& 30
						& $+7$
							&33	&39	&7.9			&$4.8^{+0.8}_{-0.7}$	&$-0.04\pm0.14$    \\
~~\fetappppip	&$ 76^{+16}_{-15}$	
					& 32
						& $+6$
					   		&24	&23	&5.6   			&$5.6^{+1.3}_{-1.2}$	&$-0.32\pm0.20$    \\
\bma{\fetapip}	&                 	
					& 
						& 
							&  	&  	&\bma{\setapip}		&\bma{\retapip}		&\bma{\Aetapip}    \\
~~\fetaggKp	&$116^{+21}_{-19}$	
					& 29
						& $+8$
							&32	&39	&6.1			&$3.6\pm0.7$		&$-0.19\pm0.16$    \\
~~\fetapppKp	&$ 37^{+13}_{-12}$	
					& 24
						& $+5$
					   		&23	&23	&2.8   			&$2.6^{+1.1}_{-1.0}$	&$-0.22\pm0.33$    \\
\bma{\fetaKp}	&                 	
					& 
						& 
							&  	&  	&\bma{\setaKp}		&\bma{\retaKp}		&\bma{\AetaKp}    \\
~~\fetaggKz	&$17^{+9}_{-7}$		
					& 27
						& $+3$
							&28	&14	&2.3			&$1.6^{+1.0}_{-0.9}$	&    \\
~~\fetapppKz	&$ 5^{+5}_{-3}$		
					& 28
						& $+1$
							&21	& 8	&1.4			&$1.1^{+1.3}_{-0.9}$	&    \\
\bma{\fetaKz}	&			
					& 
						& 
							&  	&  	&\bma{\setaKz} 		&\bma{\retaKz}{$~(<\uletaKz)$}	&    \\
~~\fetaggomega	&$13^{+7}_{-6}$		
					& 32	
						& $+1$
							&14	&35	&2.5			&$1.1^{+0.6}_{-0.5}$	&    			\\
~~\fetapppomega	&$ 2^{+7}_{-5}$		
					& 6
						& $-1$
							&11	&20	&0.6			&$0.6^{+1.3}_{-1.0}$	&			\\
\bma{\fetaomega}&			
					& 
						& 
							&  	&  	&\bma{\setaomega} 	&\bma{\retaomega}{$~(<\uletaomega)$}	&   \\
~~\fetaggrhop	&$126^{+34}_{-32}$	
					& 12
						& $+18$
							&16	&39	&3.7   			&$ 7.3^{+2.4}_{-2.2}$	&$0.10\pm0.23$    \\
~~\fetappprhop	&$ 65^{+22}_{-20}$	
					& 15
						& $+3$
							&11	&23	&3.4			&$10.6^{+3.7}_{-3.5}$	&$-0.14\pm0.31$    \\
\bma{\fetarhop}	&                 	
					& 
						& 
							&  	&  	&\bma{\setarhop}	&\bma{\retarhop}	&\bma{\Aetarhop}    \\
~~\fetapepppip	&  $69^{+13}_{-12}$	
					& 42
						& $+9$
							&27	&18	&5.6			&$5.5^{+1.2}_{-1.1}$	&$0.09\pm0.18$    \\
~~\fetaprgpip	&  $30^{+16}_{-15}$	
					& 13
						& $+9$
					   		&17	&30	&1.4   			&$1.8^{+1.3}_{-1.2}$	&$0.58\pm0.44$    \\
\bma{\fetappip}	&                 	
					& 
						& 
							&  	&  	&\bma{\setappip}	&\bma{\retappip}	&\bma{\Aetappip}    \\
\dbline
\end{tabular}
\end{table*}

The branching fraction for each decay is obtained from the measured
yield, corrected for the fit bias and for the selection efficiency, and the number of \BB\ pairs.
We assume equal decay rates of the \UfourS\ to \BpBm\ and \BzBzb .
In Table~\ref{tab:results} we show for each decay mode the measured
branching fraction together with the event yield and efficiency, and \acp\ when applicable.
The purity is the ratio of the signal yield ($Y_S$) to the effective background
plus signal ($Y^{\rm eff}_B+Y_S$), which we estimate as the square of the
uncertainty in the signal yield ($Y^{\rm eff}_B+Y_S\equiv\sigma^2_{Y_S}$).

The statistical uncertainties in the signal yield and \acp\ are taken as the
change in the central value when the quantity $-2\ln{\cal L}$ increases
by one unit from its minimum value. The significance is taken as the
square root of the difference between the value of $-2\ln{\cal L}$ (with
systematic uncertainties included) for zero signal and the value at its
minimum.


For each mode the
measurements for separate daughter decays are
combined by adding the values of $-2\ln{\cal L}$ as functions of
branching fraction, taking proper account
of the correlated and uncorrelated systematic uncertainties described below~\cite{PRD}.
For \fetaomega\ and \fetaKz\ we quote 90\% confidence level (C.L.) 
upper limits, taken to be the branching fraction below which lies 90\% of
the total of the likelihood integral in the positive branching fraction
region.

\begin{figure}[!htb]
 \includegraphics[angle=0,scale=0.4]{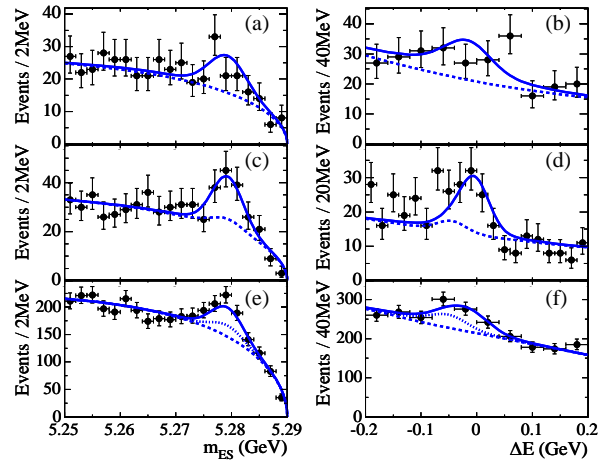}
\vspace{-0.2cm}
 \caption{\label{fig:projdeMb}
 The $B$ candidate \mb\ (left) and \DE\ (right) projections obtained with
a cut on the signal likelihood (see text) for \etarhop\ (a, b),
 \etappip\ (c, d), and combined \etapip\ and \etaKp\ (e, f).
 Points with uncertainties represent the data, solid curves the full fit functions,
 dashed curves the background functions and the dotted curves the background plus signal \fetaKp\ functions.}
\end{figure}

In Fig.\ \ref{fig:projdeMb} we show projections onto \mes\ and \DE\ 
of subsamples enriched with a mode-dependent threshold requirement on
the signal likelihood (computed without of the variable plotted) that optimizes the sensitivity.

Most of the
uncertainties arising from lack of knowledge of the
PDFs have been included in the statistical uncertainty since most background
parameters are free in the fit.  For the signal the uncertainties in PDF
parameters are estimated from the consistency of fits to MC and data in
control modes with similar final states.  Varying the signal PDF parameters within these errors,
we estimate the mode-dependent uncertainties due to the signal PDFs to be 1--8 events.
We verify the validity of the fit procedure and
PDF shapes by demonstrating that the likelihood for each fit is
consistent with the distribution found in simulation.

The uncertainty in the fit bias correction is taken to be half of the
correction itself.  Similarly we estimate the uncertainty from modeling
the \BB\ backgrounds by taking half of the
difference between the signal yield fitted with and without the \BB\
background component.

Uncertainties in our knowledge of the reconstruction efficiency, found from auxiliary 
studies on inclusive control samples~\cite{PRD}, include $0.6\%$ per primary track, $0.8\%$ per track
from a resonance, $1.5\%$ per photon, and 2.1\%\ for a \KS.
Our estimate of the systematic uncertainty in the number of \BB\ pairs is 1.1\%.
Published data \cite{PDG2004}\ provide the
uncertainties in the $B$-daughter product branching fractions (1--3\%).
The uncertainties in the efficiency of the event selection
are 1\% (4\%\ in \etaprgpip) for the requirement on
\costhr\ and 1\% for PID.  Using several large inclusive kaon and
$B$-decay samples, we find a systematic uncertainty for \acp\ of 1.1\%,
due mainly to the dependence of reconstruction efficiency on the charge,
for the high momentum pion from \etapip, \fetaKp and \fetappip .  The corresponding number for
the softer charged pion from the $\rhop$ in \etarhop\ is 2\%.


In this Letter, we have presented improved measurements of branching fractions
for six charmless $B$-meson decays.
All branching fractions are in agreement with theoretical predictions.
The previously unobserved \etarhop\ and
\etappip\ decay modes are seen with significance $\setarhop\sigma$ and $\setappip\sigma$,
respectively. For the charged modes, we also determine the charge asymmetries.
These are found to be consistent with zero within their uncertainties.


We are grateful for the excellent luminosity and machine conditions
provided by our \pep2\ colleagues, 
and for the substantial dedicated effort from
the computing organizations that support \babar.
We wish to acknowledge support from the University of Colorado Undergraduate
Research Opportunities Program.
The collaborating institutions wish to thank 
SLAC for its support and kind hospitality. 
This work is supported by
DOE
and NSF (USA),
NSERC (Canada),
IHEP (China),
CEA and
CNRS-IN2P3
(France),
BMBF and DFG
(Germany),
INFN (Italy),
FOM (The Netherlands),
NFR (Norway),
MIST (Russia), and
PPARC (United Kingdom). 
Individuals have received support from CONACyT (Mexico), A.~P.~Sloan Foundation, 
Research Corporation,
and Alexander von Humboldt Foundation.

\end{document}